\documentclass[twocolumn,showpacs,prb,amsmath,amssymb,floatfix]{revtex4-1}
\usepackage{amsmath,amssymb,color}
\usepackage{bm}
\usepackage{epsfig}
\usepackage{psfrag}
\usepackage[normalem]{ulem}
\newcommand{\bd}{\bm}

\begin{document}

\title{Multilogarithmic velocity renormalization in graphene}

\author{Anand Sharma and Peter Kopietz}
\affiliation{Institut f\"{u}r Theoretische Physik, Universit\"{a}t Frankfurt,  Max-von-Laue Strasse 1, 60438 Frankfurt, Germany}

\date{\today}

 \begin{abstract}

We reexamine the effect of long-range Coulomb interactions on the
quasiparticle velocity in graphene. 
Using a nonperturbative functional renormalization group approach with
partial bosonization in the forward scattering channel and momentum transfer cutoff 
scheme, we calculate the quasiparticle velocity, $v (k)$, and the quasiparticle residue, $Z$,
with frequency-dependent polarization. 
One of our most striking results is that $v ( k )  \propto   \ln [ C_k (\alpha) /  k ]$ 
where the momentum- and interaction-dependent cutoff scale $C_k (\alpha) $
vanishes logarithmically for $k \rightarrow 0$. Here $k$ is measured with respect to 
one of the charge neutrality (Dirac) points and $\alpha=2.2$ is the strength of dimensionless 
bare interaction.
Moreover, we also demonstrate that 
the so-obtained multilogarithmic singularity 
is reconcilable with the perturbative expansion of $v (k)$ in powers of the bare interaction.

\end{abstract}


\maketitle

\section{Introduction}

In the last decade, due to the exceptional physical properties~\cite{Geim07,Castroneto09,Dassarma11} of graphene, 
the two-dimensional all carbon material
has been envisaged as a natural candidate for 
various low-dimensional device applications~\cite{Schwierz10,Bonaccorso10,Avouris10}. 
But certain characteristics of electron-electron interactions in graphene~\cite{Kotov12}
still remain unsettled, as explained below, despite many sincere attempts to unravel 
the significance of interaction 
effects~\cite{Gonzalez94,Gonzalez99,Herbut06,Bostwick07,Dassarma07,Novoselov07,Mishchenko07,Vafek07,Son07,Vafek08,
Herbut08,Drut08,Foster08,Kotov08,Elias11,Sharma12,Chae12,Barnes14,Hofmann14,Bauer15}.

An important manifestation of many-body interactions in freely suspended graphene 
was observed in the quasiparticle velocity, $v (k)$, which was experimentally~\cite{Elias11} 
shown to acquire a logarithmic enhancement close to one of its charge neutrality (Dirac) points.
Such a behavior was found to be comparable to the first-order perturbation theory~\cite{Gonzalez94}, i.e., 
\begin{equation}
\frac{v (k )}{v_F} \approx 1 + \frac{\alpha}{4} \ln \left( \frac{\Lambda_0}{k} \right),
\label{eq:fopt}
\end{equation}
with $\Lambda_0$ being the ultraviolet cutoff of the order of inverse lattice spacing 
of the underlying honeycomb lattice and 
the momentum $k$ is measured relative to the Dirac point.
Here $\alpha = e^2/v_F \approx 2.2$ is the strength of dimensionless bare interaction in vacuum
with $e$ being the electron charge and $v_F$ is the bare Fermi velocity at the Dirac points.
Since $\alpha$ is of order unity, it signifies the failure of 
perturbation theory, i.e., expansion of $v(k)$ in powers of $\alpha$, 
in explaining the experimental results. 
Due to the lack of dielectric and conduction screening in freely standing and undoped graphene 
the theories based on the 
random phase approximation (RPA)~\cite{Hofmann14} remain doubtful. 
Moreover, large-$N$ approximation~\cite{Gonzalez99,Herbut06,Son07,Drut08,Foster08}, 
an expansion in the inverse number $N$ of fermionic flavors, is questionable because in the physically 
relevant case of graphene $N=4$ is rather small.
 
Recently, Barnes {\it{et al.}}~\cite{Barnes14} demonstrated the 
breakdown of perturbation theory (which, however, does not imply nonrenormalizability 
of the underlying field theory) 
and showed that the 
direct expansion of $v(k)$ in powers of $\alpha$ generates a series 
involving all powers of logarithms,
\begin{eqnarray}
 \frac{ v (k)}{v_F} & = & 1 + \sum_{n=1}^{\infty} F_n ( \alpha )
 \left[ \ln \left( \frac{\Lambda_0}{k} \right) \right]^n ,
 \label{eq:vpert}
\end{eqnarray}
where the interaction-dependent coefficients have the following 
expansion in powers of $\alpha$ :
 \begin{eqnarray}
 F_1 ( \alpha ) & = &  f_1^{(1)} \alpha + f_1^{(2)} \alpha^2 + f_1^{(3)} \alpha^3 + {\cal{O}} ( \alpha^4 ),
 \label{eq:expcoeff1} 
 \\
 F_n ( \alpha ) & = & f_{n}^{(n+1)} \alpha^{n+1} + {\cal{O}} ( \alpha^{n+2} ),
 \mbox{  for  $n \geq 2 $}.
 \label{eq:expcoeffn}
 \end{eqnarray}
The superscripts correspond to the powers of $\alpha$ and hence to the 
number of loops in the corresponding Feynman diagrams.
The authors of Ref.~[\onlinecite{Barnes14}] also pointed out that
to order $\alpha^n$, $n \geq 2$,  the perturbation series of $v ( k )$ contains
all powers  $  [ \ln ( \Lambda_0 / k )]^m$ of the basic logarithm  in the range
$m =1, \ldots, n-1$. Thus, from three-loop order onwards, 
the higher logarithmic powers start appearing in the 
perturbative expansion. 
The numerical value of the one-loop coefficient is known to be $f_1^{(1)} = 1/4$
but for the two-loop coefficient $f^{(2)}_1$ 
there exist conflicting results~\cite{Mishchenko07,Vafek08,Sharma12,Barnes14}. 
It is clear, however, that the above series in powers of logarithms cannot 
be resumed to a power law.
Moreover, the perturbative expansion in Eq.~(\ref{eq:vpert}) seems to be 
incompatible with previous renormalization group (RG) calculations~\cite{Gonzalez94,Bauer15}
as well as with resummation schemes based on the 
RPA~\cite{Hofmann14} which did not find higher powers 
of $\ln ( \Lambda_0 / k )$. This is a clear indication of the ambiguity  
concerning the nature of interaction effects in graphene.  
Thus it is important to understand and resolve this enigma 
before the material properties (quasiparticle velocity) 
can be engineered~\cite{Hwang12} for promising applications.

Motivated by this fact, in this work, we reexamine interaction effects on the 
quasiparticle velocity in graphene. 
We argue that the perturbative expansion in Eq.~(\ref{eq:vpert}) 
is reconcilable with the RG by showing that the higher logarithmic powers 
can be resumed with the help of non-perturbative 
functional renormalization group (FRG) flow equations~\cite{Kopietz10,Metzner12} 
to yield an expression of the form
 \begin{equation}
 \frac{ v (k)}{v_F}  = 1 + B ( \alpha ) \ln \left( \frac{C_{k} ( \alpha )}{k} \right),  
 \label{eq:vkres}
 \end{equation}
where the momentum- and interaction- dependent cutoff scale $C_k(\alpha)$ vanishes 
logarithmically for $k \rightarrow 0$. 
We believe that Eq.~(\ref{eq:vkres}) gives the true asymptotic behavior of the 
quasiparticle velocity $v(k)$ close to the Dirac points of undoped graphene.
In order to corroborate the validity of our calculation, we show that 
the direct expansion of Eq.~(\ref{eq:vkres}) in powers of $\alpha$ reproduces the 
structure of the perturbation series given in Eq.~(\ref{eq:vpert}).

The rest of this paper is organized as follows. 
In Sec.~\ref{model}, we introduce our low-energy effective model and derive a closed 
FRG flow equation for its self-energy using the momentum transfer cutoff scheme~\cite{Schuetz05}. 
In Sec.~\ref{results}, we present the results of the solution of FRG flow equations 
within static approximation as well as including dynamic screening. 
In the concluding Sec.~\ref{summary}, we summarize our findings and present an outlook.

\section{Model, method, and FRG flow equations}\label{model}

We describe the low-energy physics of graphene 
by considering an effective model consisting of fermions 
with momenta close to the Dirac points 
which interact via long-range Coulomb forces on a two-dimensional honeycomb lattice.
It is convenient to decouple the interaction
with the help of a Hubbard-Stratonovich field $\phi$, so that 
our bare Euclidean action~is
 \begin{eqnarray}
 S_{\Lambda_0} [\psi , \phi ] 
& = &  - \sum_{p \sigma }\int_K \psi^{\dagger}_{p \sigma} ( K ) [G^0_{p} 
( K )]^{-1} \psi_{p \sigma }  ( K )  \nonumber
 \\
 &+  & \frac{1}{2}
  \int_{{Q}} 
 \left[  f_{\bd{q}}^{-1}  \phi ( - {Q} )   \phi ( {Q} )  
 +  2 i
  \rho ( - Q ) \phi ( Q) \right] ,
 \hspace{7mm}
 \label{eq:Sbare}
 \end{eqnarray} 
where $\psi_{p \sigma } ( K )$ is a two-component fermion field
labeled by the Dirac point $p = \pm$, the spin projection $\sigma = \pm$,
and the frequency-momentum label
$K = (i \omega , \bd{k}  )$. Here 
$i \omega $ is a fermionic Matsubara frequency.
The two components of $\psi_{p \sigma } ( K )$
are associated with the two sublattices of the underlying
honeycomb lattice. 
The inverse fermionic propagator
is given by the following $2 \times 2$ matrix
in the sublattice labels,
\begin{equation}
[G^0_{p}  ( K )]^{-1} = i \omega - p v_F {\bd{\sigma}} \cdot {\bd{k}}  ,
\label{eq:Gbare}
\end{equation}
where the components of the two-dimensional vector
$\bd{\sigma} = [ \sigma^x , \sigma^y ]$ are the Pauli matrices
acting in sublattice space.
The bare propagator of the bosonic Hubbard-Stratonovich field $\phi ( Q )$
is given by the two-dimensional Fourier transform
$f_{\bd{q}} = 2 \pi e^2 / | \bd{q} |$ of the Coulomb interaction
and the composite field
\begin{equation}
\rho ( Q) = \sum_{p \sigma} \int_K \psi_{p \sigma}^{\dagger} ( K ) 
\psi_{p \sigma} ( K + Q ) 
\end{equation}
 represents the density.
The bosonic field $\phi ( Q )$ 
is labeled by $Q = ( i \bar{\omega} , \bd{q} )$,
where $i \bar{\omega}$ is a bosonic Matsubara frequency, and
the integration symbols are $\int_K = ( 2 \pi )^{-3} \int d \omega 
\int d^2 k$ and $\int_Q = ( 2 \pi )^{-3} \int d \bar{\omega} 
\int d^2 q$.

We now write down FRG flow equations for our low-energy theory defined by Eq.~(\ref{eq:Sbare}) using
the momentum transfer cutoff scheme proposed in Ref.~[\onlinecite{Schuetz05}].
In this scheme, we introduce a cutoff 
$\Lambda$ only in the bosonic sector, such that it
restricts the momentum  transferred by the bosonic Hubbard-Stratonovich  field
to the regime $ | \bd{q} | > \Lambda$.
For our purpose, it is sufficient to work with
a sharp cutoff which amounts to replacing the bare interaction by 
$ \Theta ( | \bd{q} | - \Lambda ) f_{\bd{q}}$.  
In systems where the interaction is dominated by small
momentum transfers
this cutoff scheme has several 
advantages~\cite{Kopietz10,Schuetz05,Schuetz06,Drukier12}.
In particular, it does not violate Ward identities related to particle number conservation.
In fact,
in Ref.~[\onlinecite{Schuetz05}] it was shown that in this 
cutoff scheme the FRG flow equations for the one-dimensional Tomonaga-Luttinger model
can be solved exactly to rederive the nonperturbative bosonization result 
for the single-particle Green's function. 
In the present context, the advantage of this cutoff scheme is that
it can be combined with a Dyson-Schwinger equation in the bosonic sector to
derive a closed FRG flow equation for the fermionic self-energy 
from which we can extract the renormalized velocity
with a rather modest numerical effort.
In contrast, if we work with a cutoff in the fermionic sector
we have to solve 
more complicated coupled integro-differential equations to obtain the
renormalized velocity~\cite{Bauer15}.

From the general hierarchy of FRG flow equations~\cite{Kopietz10},
we obtain the following exact flow equation for the
fermionic self-energy in the momentum transfer cutoff scheme,
 \begin{eqnarray}
 \partial_{\Lambda} \Sigma_{p}^{s s^{\prime}} ( K ) & = &  \sum_{ s_1 s_2 }     \int_Q
 \dot{F} ( Q) \Gamma^{s {s_1} \phi } _{p}  ( K , K-Q ; Q ) 
 \nonumber
 \\
 &  & \times  
G^{s_1 s_2 }_{ p} ( K - Q )
 \Gamma^{s_2 s^{\prime} \phi}_{p}  ( K - Q , K ; -Q ) 
 \nonumber
 \\
 & + &  \frac{1}{2} \int_Q \dot{F} ( Q) 
 \Gamma_{p}^{ s s^{\prime} \phi \phi} ( K , K ; Q , - Q ),
 \label{eq:flowself1}
 \end{eqnarray}
where  the fermionic propagator is related to the self-energy via the Dyson equation,
\begin{equation}
 {[} G_{p } (K)]^{-1} = [ G^0_{p} (K)]^{-1}  - \Sigma_{p} (K) ,
\end{equation}
which is a $2 \times 2$ matrix equation in the sublattice basis labeled by
$s , s^{\prime} \in \{ A, B \}$.
The external legs attached to the three-legged vertices
$\Gamma_p^{s s^{\prime} \phi } (K , K^{\prime} ; Q )$
correspond to the fields $\bar{\psi}^s_p ( K )$, 
$\psi^{s^{\prime}}_p ( K^{\prime})$, and $\phi ( Q )$.
Similarly, the four-legged vertex $\Gamma_{p}^{ s s^{\prime} \phi \phi} ( K , K ; Q , - Q )$
in the last line of Eq.~(\ref{eq:flowself1}) has two fermion legs
associated with $\bar{\psi}_p^{s} (K)$ and
$\psi_p^{s^{\prime}} ( K )$, and two boson legs.
In our cutoff scheme the bosonic single-scale propagator is given by, 
\begin{equation}
\dot{F} (Q) = - \frac{\delta ( | \bd{q} | - \Lambda )}{[ \frac{\Lambda}{2 \pi e^2} + \Pi (Q)]}  ,
\end{equation}
where the bosonic self-energy $\Pi (Q)$ can be identified with the
irreducible particle-hole bubble.
Now instead of writing down another FRG flow equation for
 $\Pi (Q)$, we shall follow Refs.~[\onlinecite{Schuetz05,Drukier12}]
and close the RG flow using the exact Dyson-Schwinger equation
 \begin{eqnarray}
  \Pi  ( Q ) & = & i N_s
 \sum_{s s^{\prime} } \sum_{p} \int_K
 {G}_{p}^{s s^{\prime}} ( K) 
 G^{s^{\prime} s }_{ p} ( K - Q )
 \nonumber
 \\
 & & 
  \hspace{15mm} \times \Gamma^{s^{\prime} s^{\prime} \phi } _{p}  ( K  , K- Q  , Q ) ,
 \label{eq:skeleton}
 \end{eqnarray}
where the factor $N_s =2S+1 =2$ is the spin degeneracy.

\begin{figure}[t]
\includegraphics[width=80mm]{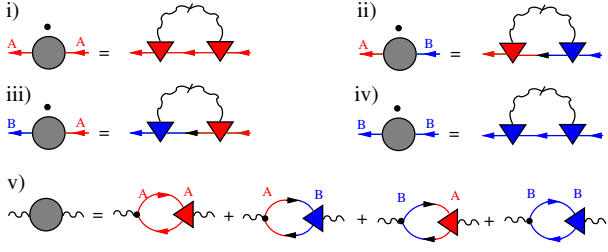}
  \caption{
The truncated FRG flow equation for the fermionic self-energy in momentum transfer cutoff scheme 
is depicted from (i) - (iv) with $A$ and $B$ being the sublattice labels. 
The dot on the left-hand side represents the derivative with respect to the cutoff. 
The arrows illustrate the exact fermionic propagators, while the bosonic single-scale propagators 
are shown by slashed wavy lines.  
The graphical representation of the exact skeleton equation for the bosonic self-energy is shown in (v). 
The triangles represent the flowing three-legged vertices, while the bare ones are shown by black dots.
}
\label{fig:diagrams}
\end{figure}

To obtain a closed system of equations, we need additional equations
of the three- and four-point vertices appearing in Eqs.~(\ref{eq:flowself1}) and
(\ref{eq:skeleton}). Our truncation strategy is based on the classification
of the vertices according to their relevance at the quantum critical point 
describing undoped graphene at vanishing temperature~\cite{Bauer15}. 
We retain only the marginal part of all vertices 
which are finite at the initial RG scale.
This implies that we should neglect the
mixed four-point vertex
$\Gamma_{p}^{ s s^{\prime} \phi \phi} ( K , K ; Q , - Q )$ (which is irrelevant)
and the sublattice changing three-point vertices corresponding to the
field combinations $\bar{\psi}^A \psi^B \phi$ and
$\bar{\psi}^B \psi^A \phi$ (which vanish at the initial scale).
Moreover, the momentum- and frequency-dependent part of the
three-point vertices is irrelevant so that
it is sufficient to retain only 
\begin{equation}
 \Gamma_p^{AA \phi } (0,0;0 ) =  \Gamma_p^{BB \phi } (0,0;0 ) = i \gamma_{\Lambda} .
\end{equation}
With the above approximations the exact FRG flow equation (\ref{eq:flowself1})
reduces to
 \begin{equation}
 \partial_{\Lambda} \Sigma_p ( K ) = - \gamma_{\Lambda}^2 \int_Q 
 \dot{F} ( Q ) G_p ( K-Q ),
 \label{eq:selftrunc}
 \end{equation}
while the Dyson-Schwinger equation (\ref{eq:skeleton}) becomes 
  \begin{equation}
  \Pi  ( Q )  =  -  \gamma_{\Lambda}  N_s \sum_{s s^{\prime} }
 \sum_{p} \int_K
 {G}_{p}^{s s^{\prime}} ( K) 
 G^{s^{\prime} s }_{ p} ( K - Q )
 \label{eq:skeleton2}
 \end{equation}
as graphically depicted in Fig.~\ref{fig:diagrams}. 
Finally, the FRG flow is closed by relating the vertex $\gamma_{\Lambda}$ to the
wave-function renormalization factor $Z_{\Lambda}$ 
via the Ward identity~\cite{Bauer15,Gonzalez10}  
$\gamma_{\Lambda} = 1 / Z_{\Lambda}$, 
which can be derived by comparing the FRG flow equation for
$\gamma_{\Lambda}$ with the flow equation for $Z_{\Lambda}$.

We determine the cutoff-dependent quasiparticle residue, $Z_{\Lambda}$, and  
quasiparticle velocity, $v_\Lambda$, by 
expanding the self-energy for small frequencies and momenta,
 \begin{equation}
  \Sigma_{p} ( K ) = ( 1 - Z_{\Lambda}^{-1} ) i \omega
 - ( 1 - Y_{\Lambda}^{-1} ) p v_F \bd{\sigma} \cdot \bd{k} 
 + \ldots,
 \label{eq:sigmalow}
 \end{equation}
so that the fermionic propagator is 
 \begin{equation}
 G_{p } ( K ) = - Z_{\Lambda} \frac{ i \omega +   p v_{\Lambda}
 \bd{\sigma} \cdot \bd{k} }{ \omega^2 +  {v}_\Lambda^2
 \bd{k}^2 },
 \label{eq:Glow}
 \end{equation}
 with renormalized quasiparticle velocity
 \begin{equation}
 {v}_{\Lambda} = Z_{\Lambda} Y_{\Lambda}^{-1} v_F.
 \end{equation}
From the self-energy expression, Eq.~(\ref{eq:sigmalow}), it is clear that the RG flow of $Z_{\Lambda}$ and $Y_{\Lambda}$
can be expressed in terms of the cutoff derivative of the self-energy
as 
\begin{eqnarray}
 \Lambda \partial_{\Lambda} Z_{\Lambda} & = & \eta_{\Lambda} Z_{\Lambda},
 \\
\Lambda \partial_{\Lambda} Y_{\Lambda} & = & \tilde{\eta}_{\Lambda} Y_{\Lambda},
 \end{eqnarray}
with
 \begin{eqnarray}
\eta_{\Lambda} \ & = & \Lambda Z_{\Lambda} \lim_{ \omega \rightarrow 0}
 \frac{ \partial}{\partial ( i \omega )} \partial_{\Lambda} \Sigma^{s s}_{p } ( 0 , i \omega ),
 \\
    ( \bd{\sigma} \cdot {\hat{\bd{k}}} )  \tilde{\eta}_{\Lambda}   & = & - \Lambda Y_{\Lambda} \lim_{ | \bd{k} |  \rightarrow 0}
 \frac{ \partial}{\partial ( p v_F | \bd{k} | )} 
 \partial_{\Lambda} \Sigma^{s s^{\prime}}_{p } ( \bd{k} , 0 ).
 \end{eqnarray}
The RG flow of the renormalized velocity
${v}_{\Lambda} = Z_{\Lambda} Y^{-1}_{\Lambda}  v_F$ is therefore 
 \begin{equation}
  \Lambda \partial_{\Lambda} {v}_{\Lambda}  =   ({\eta}_{\Lambda} 
 - \tilde{\eta}_{\Lambda} ) {v}_{\Lambda}.
 \end{equation}
On substituting the low-energy form of the Green's function, Eq.~(\ref{eq:Glow}),
into the Dyson-Schwinger equation, Eq.(\ref{eq:skeleton2}), we get for 
the renormalized polarization
 \begin{eqnarray}
  \Pi ( Q) & = & \frac{ N_s}{8}
 \frac{ \gamma_{\Lambda} Z_{\Lambda}^2    \bd{q}^2}{ \sqrt{  {v}_{ \Lambda}^2  \bd{q}^2 + \bar{\omega}^2  }}.
 \label{eq:Pires}
 \end{eqnarray}
Now using the truncated FRG flow equation, Eq.~(\ref{eq:selftrunc}), 
with corresponding renormalized fermionic propagator, Eq.~(\ref{eq:Glow}), and 
renormalized polarization, Eq.~(\ref{eq:Pires}) which is used in the 
bosonic single-scale propagator, we obtain 
 \begin{eqnarray}
 \eta_{\Lambda} & = & Z_{\Lambda}^2 \gamma_{\Lambda}^2 \Lambda
 \int_Q
 \frac{ \delta ( | \bd{q} | - \Lambda )}{ 
 \frac{ \Lambda}{ 2 \pi e^2} + \Pi ( Q ) }
\frac{ \bar{\omega}^2 - ( {v}_{\Lambda} | \bd{q} | )^2 }{
 [  \bar{\omega}^2 + ( {v}_{\Lambda} | \bd{q} | )^2 ]^2 },
 \label{eq:inteta}
 \\
 \tilde{\eta}_{\Lambda} & = & Z_{\Lambda}^2 \gamma_{\Lambda}^2 \Lambda
 \int_Q
 \frac{ \delta ( | \bd{q} | - \Lambda )}{ 
 \frac{ \Lambda}{ 2 \pi e^2} + \Pi ( Q ) }
\frac{ \bar{\omega}^2  }{
 [  \bar{\omega}^2 + ( {v}_{\Lambda} | \bd{q} | )^2 ]^2 }.
 \label{eq:intetabar}
 \end{eqnarray}
Note that at zero temperature these integrations can be performed exactly.
Using the Ward identity, $\gamma_{\Lambda} = 1/ Z_{\Lambda}$,
we finally obtain
 \begin{eqnarray}
 \eta_{\Lambda} & = &
 \frac{ e^2}{{v}_{\Lambda}} 
 \int_0^{\infty} \frac{ d \epsilon}{\pi}
 \frac{1}{ u_{\Lambda} +  \sqrt{ 1 + \epsilon^2}  }
   \frac{ \epsilon^2 - 1 }{ [  \epsilon^2 +  1 ]^{3/2} }
 \nonumber
 \\
  & = & 
 \frac{ e^2}{{v}_{\Lambda}} I_1 ( u_{\Lambda} ),
 \label{eq:intetaf}
 \\
 \tilde{\eta}_{\Lambda} & = &  
 \frac{ e^2}{{v}_{\Lambda}} 
 \int_0^{\infty} \frac{ d \epsilon}{\pi}
 \frac{1}{ u_{\Lambda} +  \sqrt{ 1 + \epsilon^2}  }
   \frac{ \epsilon^2  }{ [  \epsilon^2 +  1 ]^{3/2} }
 \nonumber
 \\
  & = &
 \frac{ e^2}{{v}_{\Lambda}} I_2 ( u_{\Lambda} ),
 \label{eq:intetabarf}
 \end{eqnarray}
where we have introduced the dimensionless coupling
 \begin{equation}
 u_{\Lambda} =  Z_{\Lambda}^2 \gamma_{\Lambda} \frac{N_s}{8} \frac{ 2 \pi e^2}{ 
 {v}_{\Lambda} } = \pi N_s  Y_{\Lambda}  \frac{\alpha}{4}.
 \end{equation}
Thus, we obtain a closed system of flow equations for $Z_{\Lambda}$ and $v_{\Lambda}$. 
On introducing the logarithmic flow parameter $l = \ln ( \Lambda_0 / \Lambda )$, the 
dimensionless velocity $\tilde{v}_l = v_{ \Lambda} / v_F$, 
and on writing $Z_l = Z_{\Lambda_0 e^{-l } }$, 
we finally obtain 
 \begin{eqnarray}
\partial_l Z_l & = & - \alpha \frac{Z_l}{\tilde{v}_l} I_1  \left( c \frac{ Z_l }{ \tilde{v}_l} \right)   ,
 \label{eq:Zflow}
 \\
 \partial_l \tilde{v}_l & = &  \alpha \left[ I_2 \left(c \frac{ Z_l }{ \tilde{v}_l} \right) - 
 I_1  \left( c \frac{ Z_l }{ \tilde{v}_l} \right)   \right],
 \label{eq:vflow}
\end{eqnarray}
where 
\begin{equation}
c = \frac{\pi N_s \alpha}{4} = \frac{\pi \alpha}{2} 
\end{equation}
and the integrals $ I_{1} (u)$ as well as $ I_{2} (u)$, as given in Eqs.(\ref{eq:intetaf}) and (\ref{eq:intetabarf}), 
can be performed analytically. For $u \leq 1$, they are given by 
 \begin{eqnarray}
 I_1 (u)  & = & \frac{1}{\pi u^2}
 \left[ \pi  - 2 u- 
  \frac{ 2 - u^2 }{\sqrt{ 1 - u^2 }} \arctan \left( 
 \frac{ \sqrt{1-u^2}}{u} \right) \right],
 \nonumber \\
 \\
 I_2 ( u ) & = & \frac{1}{\pi u^2}
 \left[ \frac{\pi}{2}  -u -   \sqrt{ 1 - u^2 }
\arctan \left( 
 \frac{ \sqrt{1-u^2}}{u} \right) \right],
  \nonumber \\
 \end{eqnarray}
while for $ u > 1$ they become
\begin{eqnarray}
 I_1  ( u ) & = & \frac{1}{\pi u^2}
 \left[  \pi  - 2 u- 
  \frac{ 2 - u^2 }{2 \sqrt{  u^2 -1}} \ln \left( 
 \frac{ u + \sqrt{u^2-1}}{ u - \sqrt{u^2-1}}  \right) 
 \right],
 \nonumber \\
 \\
 I_2 ( u ) & = & \frac{1}{\pi u^2}
 \left[ \frac{\pi}{2}  -u  +   \frac{\sqrt{  u^2 -1 }}{2}
   \ln \left( 
 \frac{ u + \sqrt{u^2-1}}{ u - \sqrt{u^2-1}}  \right)      \right].
  \nonumber \\
 \end{eqnarray}
 For $u \ll 1$ the integrals can be approximated by
  \begin{eqnarray}
  I_1 (u ) & = & \frac{u}{3 \pi} - \frac{u^2}{8} + {\cal{O}} ( u^3),
 \\
 I_2 ( u ) & = & \frac{1}{4} - \frac{u}{3 \pi} + \frac{u^2}{16} + {\cal{O}} ( u^3).
 \end{eqnarray}

\section{Results}\label{results}

\subsection{Static screening approximation}

Before presenting the numerical solution of Eqs.~(\ref{eq:Zflow}) and (\ref{eq:vflow}),
it is instructive to consider the corresponding RG flow
in the approximation where the frequency dependence of the
polarization is neglected.
Approximating $\Pi ( i \bar{\omega} , \bd{q}) \approx 
\Pi ( 0, \bd{q} )$ in Eq.~(\ref{eq:Pires}),
the integral for $\tilde{\eta}_{\Lambda}$ in 
Eq.~(\ref{eq:intetabarf}) simplifies to
\begin{equation}
\tilde{\eta}_{\Lambda} = \frac{e^2}{4} 
\frac{1}{v_{\Lambda} + c  v_F}.
\end{equation}
The flow of the dimensionless velocity $\tilde{v}_l = v_{\Lambda} / v_F$,  
in this approximation, is determined by
\begin{equation}
\partial_l \tilde{v}_l =   \frac{b}{ 1 + c / \tilde{v}_l},
\label{eq:vldef}
\end{equation}
where we have defined
\begin{equation}
b = \alpha I_2 (0) = \frac{\alpha}{4} = \frac{ e^2}{4 v_F }.
\end{equation}
The solution of the differential equation (\ref{eq:vldef}) with 
initial condition $\tilde{v}_{0} =1$
is given by the solution of the implicit equation,
\begin{equation}
\tilde{v}_l   + c \ln \tilde{v}_l     = 1 + b l    ,
\label{eq:uimplicit}
\end{equation}
which can be expressed in 
terms of the so-called Lambert $W$-function~\cite{Corless96} 
$W ( x )$ (also called product logarithm),  
\begin{equation}
 \tilde{v}_l = c W \left( \frac{e^{ ( 1 + b l )/c } }{c} \right) = 
 c W
 \left[ \frac{e^{1/c}}{c}  \left( \frac{\Lambda_0}{\Lambda} 
 \right)^{\frac{b}{c} } \right] .
\end{equation}
We use the fact that by definition the Lambert $W$-function is the solution 
of  $W e^W = x$ and hence
$W ( x ) = \ln [ x / W (x)]$. 
Therefore, we may alternatively write the solution of 
the differential equation (\ref{eq:vldef}) as, 
\begin{eqnarray}
 \tilde{v}_l & = & c \ln \left[ \frac{ e^{ ( 1 + b l )/ c }}{ c 
 W \left( \frac{e^{ ( 1 + b l )/ c } }{c} \right) } \right]
 \nonumber 
 \\
 & = & 
 1 + b l  - c \ln 
 \left[ c  W \left( \frac{e^{ ( 1 + b l )/ c } }{c} \right)  \right].
\end{eqnarray}
Using the fact that $c W ( e^{1/c}/c ) =1$, we 
immediately see that our solution indeed satisfies 
the initial condition $\tilde{v}_{l=0} =1$.
Finally, in order to obtain the momentum dependence of the
quasiparticle velocity, we identify that 
$v(k) = v_{ \Lambda = k }$.
Recently, we have explicitly confirmed the validity of this identification 
using the FRG method~\cite{Bauer15}. 
Physically this procedure is based on the fact that for $ \Lambda \ll k$ the external
momentum $k$ acts as an infrared cutoff. 
Therefore, we obtain
$B ( \alpha ) = \alpha /4$ as the prefactor of the
logarithm in Eq.~(\ref{eq:vkres}).
We substitute $l = \ln ( \Lambda_0 / \Lambda )$ and obtain 
for the cutoff-dependent velocity,
\begin{equation}
\frac{ v_{\Lambda}}{v_F}  = 1 + \frac{\alpha}{4} \ln \left( \frac{C_{\Lambda} ( \alpha )}{\Lambda} \right),
 \label{eq:vvres}
\end{equation}
with scale- and interaction-dependent cutoff
 \begin{equation}
 C_{\Lambda} ( \alpha ) = \frac{\Lambda_0}{ 
 \left\lbrace c W  \left[ \frac{e^{1/c } }{c} \left( \frac{\Lambda_0}{\Lambda} 
 \right)^{b/c}
 \right] \right\rbrace^{c/b} }.
 \label{eq:Cstat}
\end{equation}
For large $x$ the Lambert $W$-function can be approximated by
\begin{equation}
 W(x) \approx \ln x - \ln \ln x + \frac{ \ln \ln x }{\ln x }.
\end{equation}
We retain only the first term of the large $x$ asymptotic expansion 
as well as using $c/b = \pi N_s $; 
we obtain for $\Lambda \rightarrow 0 $,
\begin{equation}
 C_{\Lambda} ( \alpha ) \approx \frac{\Lambda_0}{ 
\left[ 1  - c \ln c + \frac{\alpha}{4} \ln ( \Lambda_0  / \Lambda )   \right]^{\pi N_s} }.
\label{eq:Cstatasym}
\end{equation}
On substituting this approximation into Eq.~(\ref{eq:vvres}) and formally expanding the result
for small $\alpha$, we obtain 
\begin{eqnarray}
v_{\Lambda} /v_F & = & F_0 ( \alpha ) + F_1 ( \alpha ) \ln ( \Lambda_0 / \Lambda )
 + F_2 ( \alpha ) \ln^2 ( \Lambda_0 / \Lambda )
 \nonumber
 \\
 & & + {\cal{O}} [ \alpha^4 \ln^3  ( \Lambda_0 / \Lambda ) ],
\end{eqnarray}
with interaction-dependent coefficients 
 \begin{subequations}
 \begin{eqnarray}
  F_0 ( \alpha ) & = & 1 + \frac{ (\pi N_s )^2}{16} \alpha^2 \ln \left( \frac{\pi N_s }{4} \alpha \right)
 \nonumber
 \\
 & + & \frac{ ( \pi N_s )^2}{128} \alpha^3 \ln^2 \left( \frac{\pi N_s }{4} \alpha \right)
 + {\cal{O}} ( \alpha^4 ),
 \\
 F_1 ( \alpha ) & = & \frac{\alpha}{4} - \frac{ \pi N_s}{16} \alpha^2 
 \nonumber
 \\ 
  & - & \frac{ ( \pi N_s )^2}{64} \alpha^3 \ln \left( \frac{\pi N_s }{4} \alpha \right)
  + {\cal{O}} ( \alpha^4 ), 
 \\
 F_2 ( \alpha ) & = &\frac{ \pi N_s }{128} \alpha^3 + {\cal{O}} ( \alpha^4 ).  
 \end{eqnarray}
 \end{subequations}
Comparing the above results with the perturbative expansions, 
as given in Eqs.~(\ref{eq:expcoeff1}) and (\ref{eq:expcoeffn}), and
setting now explicitly $N_s =2$, we conclude that,  
within our truncation scheme,
the first two coefficients in the expansion of $F_1 ( \alpha )$ are given by
 \begin{equation}
 f_1^{(1)} = \frac{1}{4}, \; \; \; 
 f_1^{(2)}  = - \frac{\pi}{8} \approx - 0.39,
 \end{equation}
 while the coefficient of the leading $\alpha^3$ term in the weak-coupling expansion of
 $F_2 ( \alpha )$ is
 \begin{equation}
 f_2^{(3)}  =  \frac{ \pi}{64} \approx 0.049.
 \end{equation}
Keeping in mind that the coefficient $F_0 ( \alpha )$ can be
normalized to unity by redefining the ultraviolet cutoff,
we conclude that the above structure of the perturbation series
is equivalent with the series in Eq.~(\ref{eq:vpert}) 
previously derived by Barnes {\it{et al.}}~[\onlinecite{Barnes14}].

We note that according to Mishchenko~\cite{Mishchenko07}
the numerical value of the two-loop coefficient 
is $f_1^{(2)} = - \frac{5}{6} + \ln 2 \approx -0.140$.
On the other hand, 
Vafek and Case~\cite{Vafek08} found $f_1^{(2)} = -\frac{1}{3} +\frac{103}{96} - 
\frac{3}{2} \ln 2 \approx -0.300$, while
Sharma {\it{et al.}}~\cite{Sharma12} obtained $f_1^{(2)} = - \frac{1}{3}$, 
and more recently Barnes {\it{et al.}}~\cite{Barnes14} got 
$f_1^{(2)} = - \frac{2}{3} + \frac{1}{2} \ln 2 \approx - 0.320$.
Given the simplicity of our truncation, our result for the two-loop coefficient
$f_1^{(2)}  \approx - 0.39$ is reasonably close to the
results of aforementioned calculations~\cite{Mishchenko07,Vafek08,Sharma12,Barnes14}.
Note that recently Barnes {\it{et al.}}~\cite{Barnes14} found that the three-loop coefficient
$f_2^{(3)}$ is minus one-eighth of the two-loop coefficient
$f_1^{(2)}$, which is confirmed by our calculation.
Although the static screening approximation 
is not expected to give a quantitatively accurate result,
the fact that the perturbative expansion of the renormalized velocity, 
Eq.~(\ref{eq:vvres}),
in powers of $\alpha$ reproduces the known structure of
perturbation theory~\cite{Barnes14} gives us confidence that our RG approach
indeed resums the entire perturbation series in a sensible way.

\subsection{Including dynamic screening}

\begin{figure}[t]
  \centering
  \includegraphics[width=80mm]{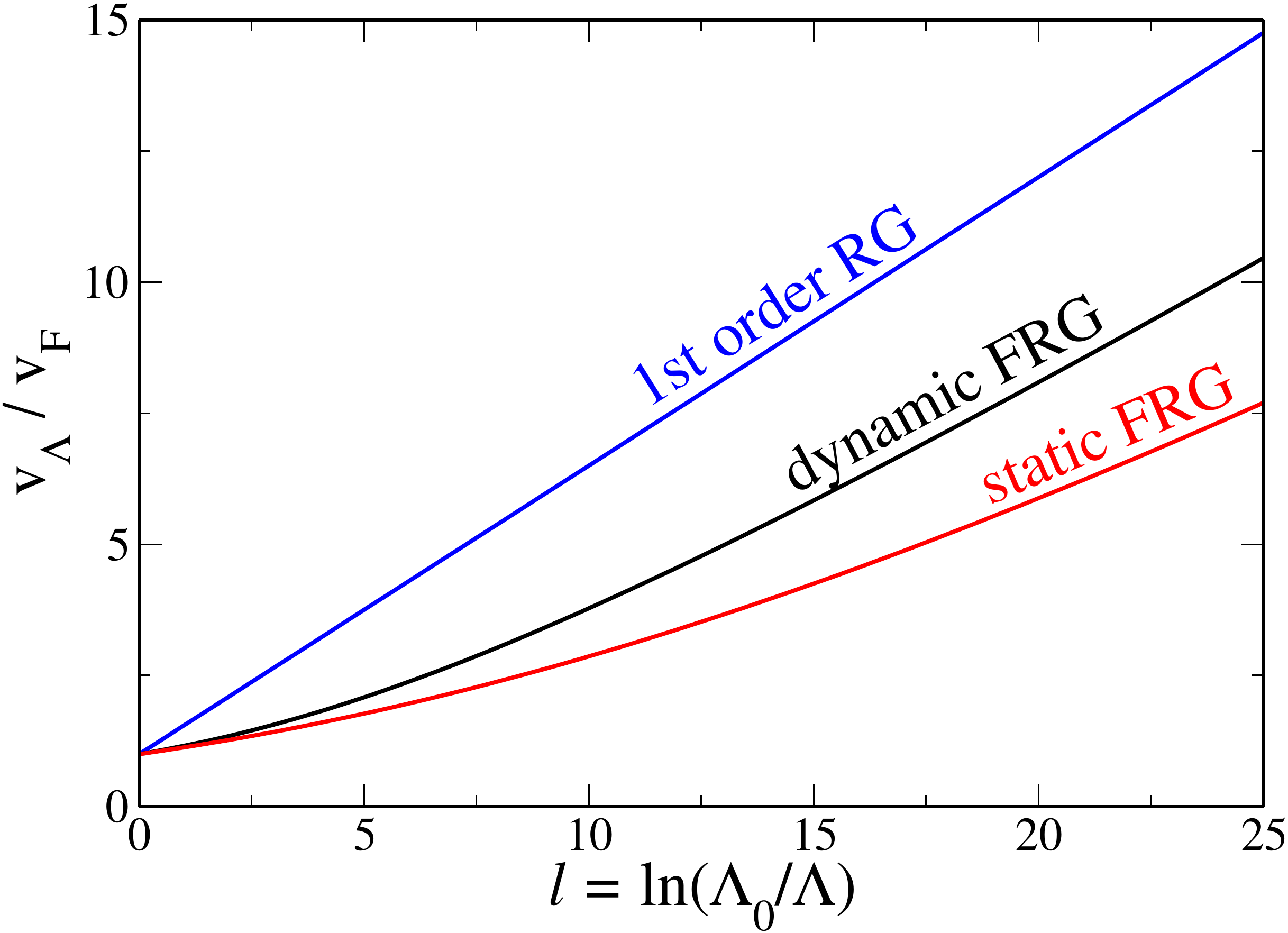}
  \caption{%
The RG flow of the dimensionless velocity $\tilde{v}_l = v_{\Lambda} / v_F$
(black middle line), 
as obtained from the numerical solution
of  Eqs.~(\ref{eq:Zflow}) and (\ref{eq:vflow}), 
is shown as a function of the flow parameter
$l = \ln ( \Lambda_0 / \Lambda )$
for $\alpha =2.2$.
The corresponding result (\ref{eq:vvres}) 
in static screening approximation
(lower red line), and the perturbative
one-loop RG result $\tilde{v}_l = 1 + ( \alpha/4 ) l$ (upper blue line) 
are also presented.
}
 \label{fig:vflow}
\end{figure}

\begin{figure}[!ht]
  \centering
  \includegraphics[width=80mm]{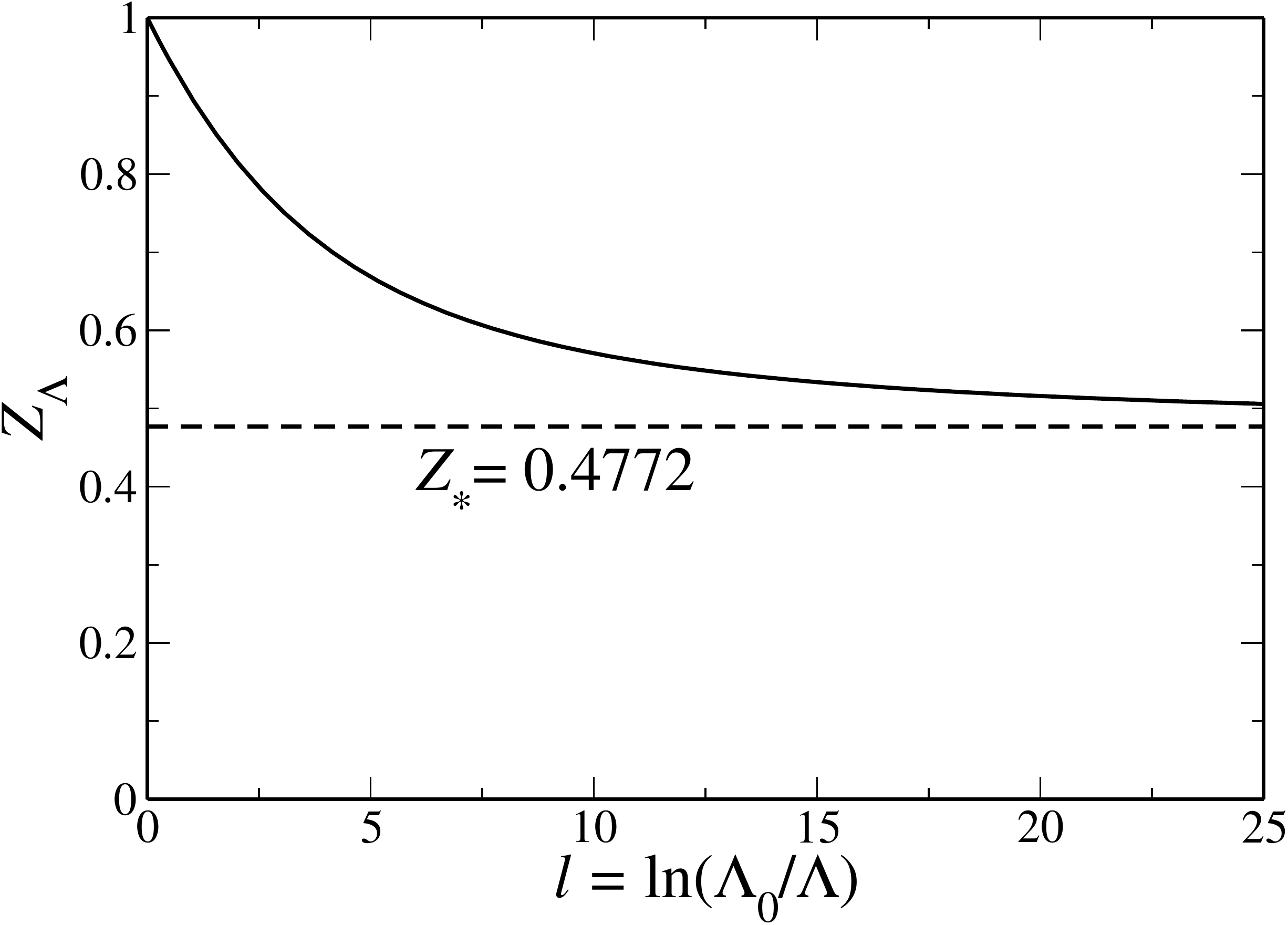}
  \caption{%
The RG flow of quasiparticle residue $Z_\Lambda$ (solid line), 
as obtained from the numerical solution
of Eqs.~(\ref{eq:Zflow}) and (\ref{eq:vflow}), 
is shown as a function of the flow parameter
$l = \ln ( \Lambda_0 / \Lambda )$
for $\alpha =2.2$.
The dashed line represents $Z_{\ast} = \lim_{l \rightarrow \infty } Z_l \approx 0.4772$. 
}
 \label{fig:Zflow}
\end{figure}

\begin{figure}[t]
\includegraphics[width=80mm]{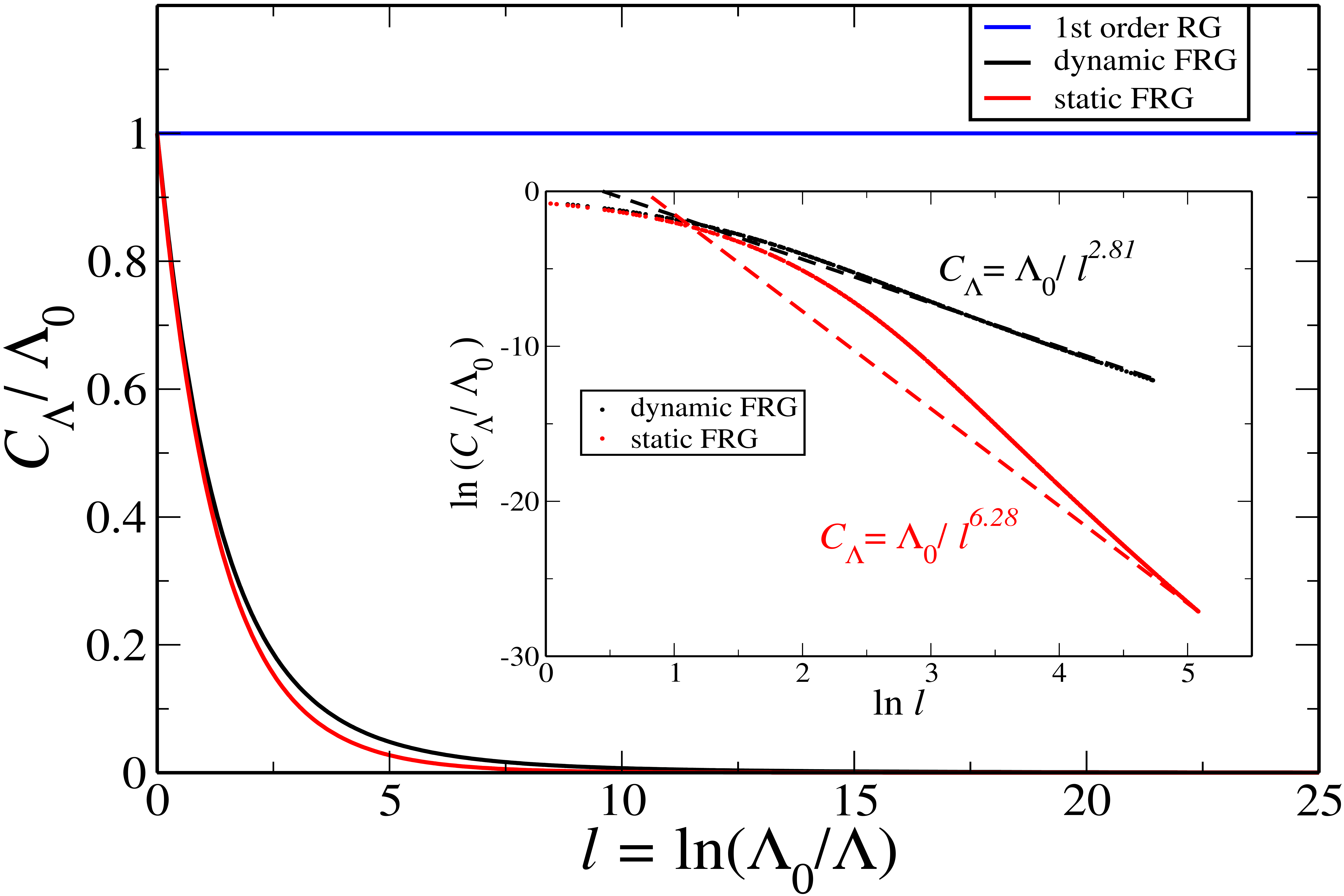}
  \caption{%
Cutoff $C_{\Lambda} ( \alpha)$
as a function of the
logarithmic flow parameter $l = \ln ( \Lambda_0 / \Lambda )$ 
for $\alpha =2.2$. The black solid line is obtained by inserting 
the solution of the flow equations (\ref{eq:Zflow}) and (\ref{eq:vflow})
into $  C_{\Lambda} ( \alpha ) = \Lambda e^{ 4 ( \tilde{v}_l -1 ) / \alpha } $.
The red solid line is the corresponding
function in static screening approximation given in
Eq.~(\ref{eq:Cstat}), while the horizontal blue solid line
is the perturbative one-loop result $C_{\Lambda} = \Lambda_0$.
The double-logarithmic plot in the inset  shows
$\ln ( C_{\Lambda} / \Lambda_0 ) $ versus
$\ln l = \ln \ln ( \Lambda_0 / \Lambda )$. The black dashed line is 
a fit of the asymptotics for large $\ln l$ given by the straight line
$\ln ( C_{\Lambda} / \Lambda_0 ) = - 2.81  \ln l$.
The red dashed line is the known asymptotics in static screening approximation,
which according to Eq.~(\ref{eq:Cstatasym}) gives
$\ln ( C_{\Lambda} / \Lambda_0 ) \sim - 2 \pi  \ln l$.
}
\label{fig:Cres}
\end{figure}

We now present our results obtained from the numerical 
solution of Eqs.~(\ref{eq:Zflow}) and (\ref{eq:vflow}) 
which take the
frequency dependence of the polarization into account.
In Fig.~\ref{fig:vflow} we show the RG flow of $\tilde{v}_l$
as a function of the logarithmic flow parameter $l = \ln ( \Lambda_0 / \Lambda )$
for the physically relevant value $\alpha = 2.2$. For comparison, we also show our
analytical result (\ref{eq:vvres}) in static screening
approximation and the perturbative
one-loop RG result $\tilde{v}_l = 1 + ( \alpha/4 ) l$.  
The corresponding RG flow of the quasiparticle residue is shown
in Fig.~\ref{fig:Zflow}. It is apparent that for $l \rightarrow \infty$ the quasiparticle residue 
approaches a finite constant, 
\begin{equation}
Z_{\ast} = \lim_{l \rightarrow \infty } Z_l \approx 0.4772 ,
\end{equation}
while the quasiparticle velocity diverges. 
To quantify this divergence, let us assume that the functional form  (\ref{eq:vvres}) obtained in static
screening approximation remains qualitatively correct so that
the cutoff-dependent velocity is of the form
\begin{equation}
 \frac{ v_{\Lambda}}{v_F}  = 1 + B ( \alpha ) \ln \left( \frac{C_{\Lambda} ( \alpha )}{\Lambda} \right),
  \label{eq:vBC}
 \end{equation}
which can be obtained from $v ( k )$ in
Eq.~(\ref{eq:vkres}) by substituting $k \rightarrow \Lambda$.
Anticipating that the cutoff function $C_{\Lambda} ( \alpha )$ vanishes logarithmically
for $\Lambda \rightarrow 0$, as seen in the static screening approximation, we may identify
\begin{eqnarray}
B ( \alpha ) & = & \lim_{ l \rightarrow \infty } \partial_l \tilde{v}_l 
 \nonumber
 \\
 & = & \alpha \lim_{ l \rightarrow \infty } 
  \left[ I_2 \left(c Z_l/ \tilde{v}_l \right) - 
 I_1  \left( c Z_l/ \tilde{v}_l \right) \right].
\end{eqnarray}
But we already know that $Z_l$ approaches a finite limit while $\tilde{v}_l$ diverges for
$l \rightarrow \infty$, so within our truncation we obtain
\begin{equation}
B ( \alpha ) = \alpha I_2 ( 0 ) = \frac{\alpha}{4} .
\end{equation}
The frequency dependence of the polarization therefore
does not modify the form (\ref{eq:vvres}) of the cutoff dependent velocity.
From the numerical solution $\tilde{v}_l$ of the flow equation (\ref{eq:vflow}), 
the cutoff function $C_{\Lambda} ( \alpha )$ in Eq.~(\ref{eq:vBC})
can be obtained as
\begin{equation}
C_{\Lambda} ( \alpha ) = \Lambda e^{ 4 ( \tilde{v}_l -1 ) / \alpha } .
\end{equation}
On inserting the perturbative one-loop result $\tilde{v}_l -  1 =  (\alpha /4 ) l
=  ( \alpha /4 ) \ln ( \Lambda_0 / \Lambda )$, 
we obtain
$C_{\Lambda} ( \alpha ) = \Lambda_0$. However, if we substitute for $\tilde{v}_l$
the solution of Eqs.~(\ref{eq:Zflow}) and (\ref{eq:vflow}) we find that for any finite
$\alpha$ the function
$C_{\Lambda} ( \alpha )$ vanishes for $\Lambda \rightarrow 0$, as shown
in Fig.~\ref{fig:Cres}.
For comparison, we also show 
the result (\ref{eq:Cstat}) of the static screening approximation.
In order to quantify the modifications due to dynamic screening, 
we present in the inset of Fig.~\ref{fig:Cres}
a double-logarithmic plot of
$\ln [ C_{\Lambda} / \Lambda_0]$ versus $\ln l = \ln \ln ( \Lambda_0 / \Lambda )$.
From the slope of the asymptotic straight line, for large $\ln l$,
we see that,
\begin{equation}
\frac{C_{\Lambda}}{\Lambda_0} \sim \frac{c_1}{[ \ln ( \frac{\Lambda_0}{\Lambda})]^{x}}   ,
\end{equation}
with $x \approx 2.81$ and $c_1$ being a numerical constant of order unity.
Note that in static screening approximation we found $x = \pi N_s = 2 \pi$,
Eq.~(\ref{eq:Cstatasym}), so that dynamic screening modifies the
power of the logarithmic decay of the cutoff  $C_{\Lambda}$ for $\Lambda \rightarrow 0 $.

\section{Summary and outlook}\label{summary}

With a strong motivation to resolve the puzzle 
related to interaction effects in graphene, we have reconsidered the problem of calculating
the renormalized quasiparticle velocity for momenta close to the
Dirac points in undoped graphene. On combining a FRG flow
equation for the fermionic self-energy with a Dyson-Schwinger equation
for the particle-hole bubble and a Ward identity for the three-legged (Yukawa) vertex,
we have derived and solved a closed system of RG flow equations for
the quasiparticle velocity and the quasiparticle residue.
In contrast to the fermionic cutoff scheme~\cite{Bauer15},
we have introduced a cutoff only in the bosonic Hubbard-Stratonovich field
which mediates the interaction in the forward scattering channel.
An important advantage of this  momentum transfer cutoff scheme~\cite{Schuetz05} is that, 
in the static limit, the flow equation for the renormalized velocity can be solved
exactly and the asymptotic behavior of the velocity can be extracted analytically.\\
\indent Our main result is that the cutoff scale below which the 
logarithmic singularity of the quasiparticle velocity becomes apparent
is itself logarithmically suppressed.
In static screening approximation, we expand our RG result in powers
of the bare coupling $\alpha$ and  
reconcile with the peculiar structure of the perturbative expansion
of $v (k)$ in powers of $\alpha$ as found by
Barnes {\it{et al.}}~\cite{Barnes14}. 
Although the higher-order logarithmic corrections become
dominant only in the close vicinity of the Dirac points,  
which probably cannot be resolved experimentally,
it is conceptually important to highlight the character 
of the multilogarithmic singularity
of the renormalized velocity and thus unfolding the nature of interaction effects in graphene.

As an outlook, our approach might also be useful 
in determining the critical interaction strength 
related to chiral symmetry breaking in graphene~\cite{Drut09}.
Recently, this problem was studied by Katanin ~\cite{Katanin16} using a purely 
fermionic FRG  approach who found that vertex corrections are crucially important
to obtain an accurate estimate of the critical interaction strength in graphene.

\section*{Acknowledgments}\label{acknowledgments}
We thank C. Bauer and A. R\"{u}ckriegel for fruitful discussions.

\end{document}